\documentclass[11pt]{article}
\usepackage{graphicx}              

\newcommand{\bc}{\begin{center}}
\newcommand{\ec}{\end{center}}

\begin{document}

\newcommand{\bu}{{\bf u}}

\begin{titlepage}
\date{}
\title{{\Large\bf Indeterminism and Time Symmetry are Incompatible: \\Reply to R\c{e}bilas
}}
\author{\Large Avshalom C. Elitzur$^{a}$, Shahar Dolev$^{b}$
}
\maketitle
\begin{center}
$^{a}$ Chemical Physics Department,\\ Weizmann Institute of Science, \\ 76100
Rehovot, Israel.\\

\noindent
E-mail:cfeli@weizmann.ac.il

\vspace{0.1cm}

$^{b}$  The Kohn Institute for the History and  Philosophy of Sciences,\\
Tel-Aviv University, 69978 Tel-Aviv, Israel.\\

\noindent
E-mail:shahar@email.com

\end{center}
\thispagestyle{empty}
\abstract\noindent
R\c{e}bilas argues that time-reversal can occur even in an indeterministic system. This hypothesis is untestable, hence lying beyond physics. 

\vspace{1cm}


\begin{center}
\end{center}
\end{titlepage}


It is common knowledge that entropy {\it decreases} towards the past. This formulation of the Second Law, although unusual, accords with the relativistic account of time as a dimension that can be read in either direction. Reading the universe's history backwards, one can say that the high-entropy states in the future constitute very remarkable arrangements of atoms that ``converge" into ordered state in the past. 

We have pointed out \cite{Eli1} an exception to this symmetry. Our simulation (Figures 1 \& 2) shows that Hawking's alleged information-loss \cite{Haw} is equivalent to the disturbance in Fig.~2, hence Hawking cannot ascribe time's arrow to initial conditions. Therefore, once determinism fails even slightly, the Second Law can be stated only in the forward direction of time. 

R\c{e}bilas \cite{Reb} objects to us by considering a system whose initial conditions are such that would lead to high entropy, yet an indeterministic event causes its entropy to decrease nonetheless. He concedes that such an event is very improbable, yet in principle may happen. 

Notice, however, that R\c{e}bilas is actually time-reversing 2b {\it together with the indeterministic event}. Here, the interference indeed appears ``indeterministic" within the system, but it is entirely deterministic for the experimenter, who must reproduce and time-reverse it with utmost precision. Consequently, one must argue that either {\it i)} all indeterministic events in our world are deterministic in some hidden level or {\it ii)} an external ``experimenter" carefully introduces the appropriate indeterministic events into our universe. The former claim is tautological for our argument, while the latter belongs to the realm of religion.

Moreover, our universe, according to Hawking, contains not one but {\it numerous} indeterministic events (For a similar claim on general relativistic grounds see \cite{Ear}). R\c{e}bilas' account must therefore run as follows: Future states are such that entropy must increase towards the {\it past}, yet a series of indeterministic events constantly tip the universe to ever decreasing entropy towards the past. Such countless coincidences that keep producing an accumulating effect amount, again, to divine supervision \cite{it}.

This holds for R\c{e}bilas' second objection, concerning our argument about the intrinsic time asymmetry (given elsewhere in greater detail \cite{Eli2}).

\begin{figure}
\begin{center}
\includegraphics[scale=0.95]{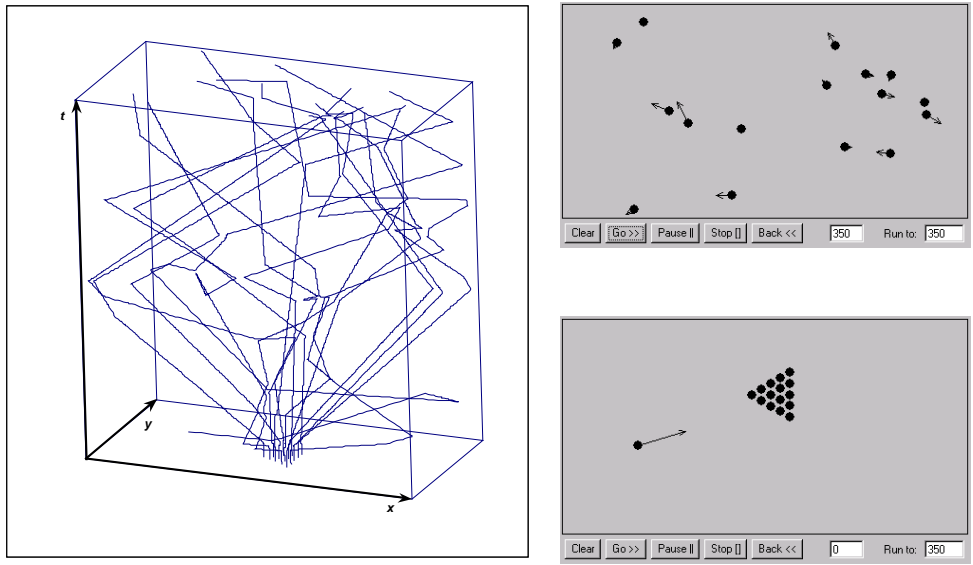}
\renewcommand{\thefigure}{1a}
\label{Fig1a}
\setlength{\abovecaptionskip}{0cm}
\setlength{\belowcaptionskip}{0.8cm}
\caption{A computer simulation of an entropy increasing process, with the initial and final states (right) and the entire process using a spacetime diagram (left). One billiard ball hits a group of ordered balls at rest, dispersing them all over the table. After repeated collisions between the balls, the energy and momentum of the first ball is nearly equally divided between the balls.}
\end{center}
\end{figure}

\begin{figure}
\begin{center}
\includegraphics[scale=0.95]{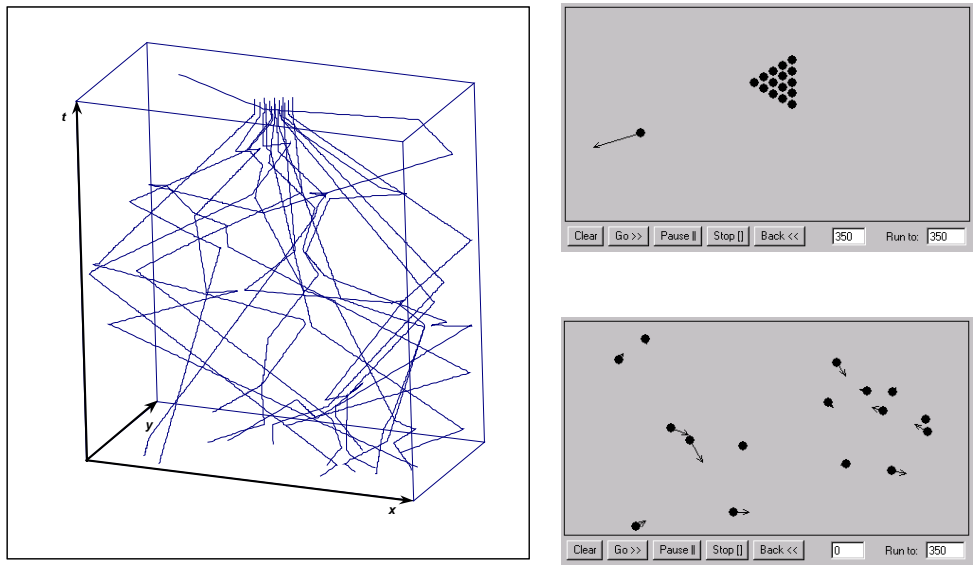}
\renewcommand{\thefigure}{1b}
\label{Fig1b}
\setlength{\abovecaptionskip}{0cm}
\setlength{\belowcaptionskip}{0cm}
\caption{The time-reversed process. All the momenta of the balls are reversed at $t_{350}$. Eventually, the initial ordered group is re-formed, as at $t_0$, ejecting back the first ball.}
\end{center}
\end{figure}

\begin{figure}[t]
	\begin{minipage}[t]{0.45\linewidth}
		\centering
		\includegraphics{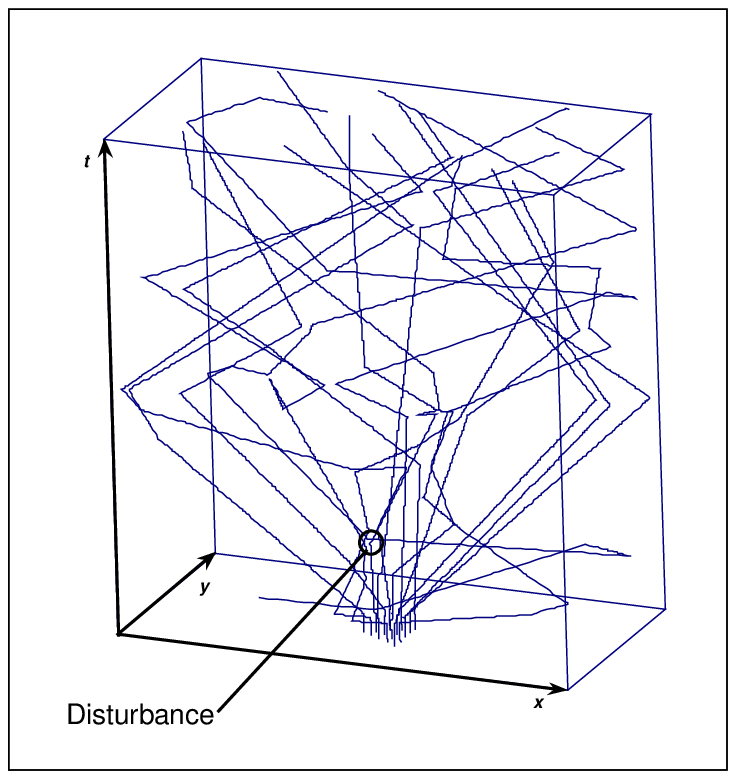}
		\renewcommand{\thefigure}{2a}
		\label{Fig2a}
		\caption{The same simulation as in 1a, with a slight disturbance in the trajectory of one ball (marked by the small circle). Entropy increase seems to be indistinguishable from that of 1a.}
	\end{minipage}%
	\hspace{0.1\linewidth}%
	\begin{minipage}[t]{0.45\linewidth}
		\centering
		\includegraphics{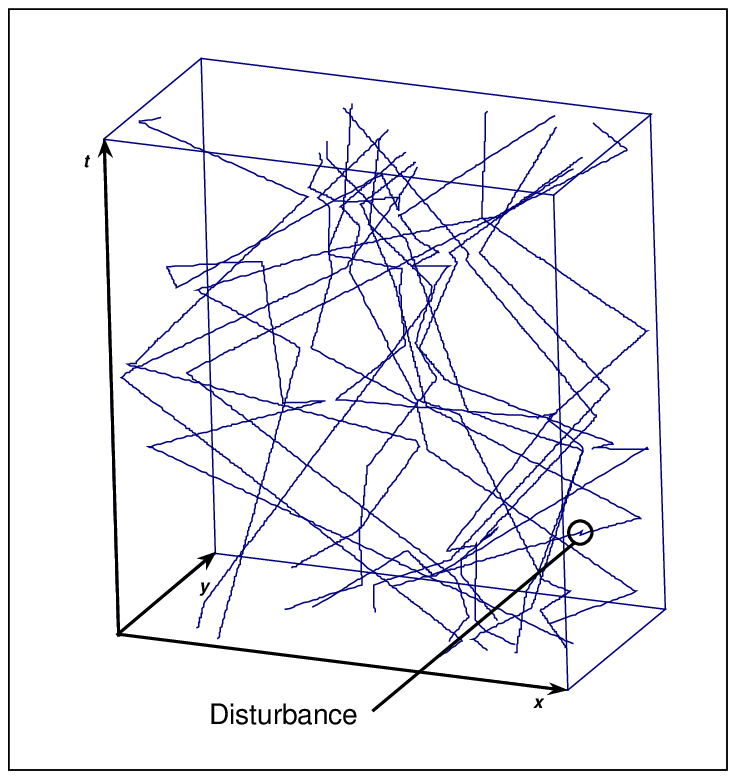}
		\renewcommand{\thefigure}{2b}
		\label{Fig2b}
		\caption{The same computer simulation as in 1b, with a similar disturbance. Here, the return to the ordered initial state fails.}
	\end{minipage}
	\vspace*{1cm}
\end{figure}

\bibliographystyle{unsrt}

\begin{thebibliography}{28}
\bibitem{it} It is one thing to ascribe the second law to the universe's boundary conditions, excusing physics from going beyond them, and another thing to introduce inexplicable coincidences into every stage of the universe's evolution!
\end{thebibliography}

\end{document}